\documentclass[a4paper, amsfonts, amssymb, amsmath, reprint, showkeys, nofootinbib, twoside,longbibliography, aps]{revtex4-1}
\usepackage[utf8]{inputenc}
\usepackage[english]{babel}
\usepackage[utf8]{inputenc}
\usepackage{printlen}
\usepackage[detect-all]{siunitx}
\usepackage{comment}
\usepackage{soul,xcolor}
\usepackage[utf8]{inputenc}
\usepackage{amsmath}
\usepackage{graphicx}
\usepackage[normalem]{ulem} 
\usepackage[colorinlistoftodos, color=green!40, prependcaption]{todonotes}

\usepackage{graphicx} 

\begin{document}
\title{Stokes microcombs in silicon nitride microresonators}

\author{Artem~E.~Shitikov\textsuperscript{1}}
\author{Alina~N.~Golodukhina\textsuperscript{1,2}}
\author{Nikita~Yu.~Dmitriev\textsuperscript{1}}
\author{Darya~M.~Sokol\textsuperscript{1,3}}
\author{Valery~E.~Lobanov\textsuperscript{1}}
\author{Igor A. Bilenko\textsuperscript{1,2}}
\author{Dmitry~A.~Chermoshentsev\textsuperscript{1,3}}
\email{d.chermoshentsev@gmail.com} 

\affiliation{\textsuperscript{1}Russian Quantum Center, Skolkovo, Moscow 121205, Russia}
\affiliation{\textsuperscript{2}Faculty of Physics, Lomonosov Moscow State University, 119991 Moscow, Russia}
\affiliation{\textsuperscript{3}Moscow Institute of Physics and Technology, Dolgoprudny, Moscow Region 141700, Russia}

\date{July 2024}

\begin{abstract}
Silicon nitride microresonators have become an ubiquitous platform for cutting-edge photonics applications. Improvement in silicon nitride fabrication techniques, providing ultra-high quality-factor values up to $10^7$, has opened up new possibilities for nonlinear effects realizations in such structures. Here we report for the first time to our knowledge on the observation of the Stokes microcombs in silicon nitride on-chip microresonators exhibiting normal group velocity dispersion. Moreover, using different pump schemes, namely, a tunable laser with an isolator and a stabilized diode laser, we demonstrate on-chip stimulated Raman frequency combs including dark-pulse Raman states. We reveal a complex interplay between Kerr and Raman nonlinearities and elaborate effective method of controllable switching between predominantly Kerr-comb and predominantly Raman-comb operation. We prove the Raman-induced platicon formation by numerical model which shows perfect agreement with experimental results. These findings are of special importance for silicon nitride photonics and provide a basis for novel photonic devices. 
\end{abstract}

\maketitle
\section{Introduction}

Silicon nitride (Si$_3$N$_4$) has become a cornerstone platform for integrated nonlinear photonics owing to its low propagation loss, negligible two-photon absorption in the telecommunications band, and CMOS compatibility \cite{Moss2013,Blumenthal2018,Xiang2022}. Advances in fabrication have delivered ultra-high-$Q$ microring resonators \cite{Puckett2021,Liu2021a} that enable low-threshold four-wave mixing and coherent optical frequency-comb generation \cite{DelHaye2007,Xuan2016}. Such devices underpin applications ranging from coherent communications and microwave photonics to spectroscopy, metrology, and astronomical calibration \cite{Marin-Palomo2017,Kudelin2024,Picqu2019,Suh2019,Obrzud2019,Feldmann2021,Sun2023,Herr2014a,Liu2018,Gaeta2019}.

In whispering gallery mode (WGM) microresonators and integrated microring resonators, Kerr nonlinearity has been the main route to coherent frequency combs. Under anomalous group velocity dispersion (GVD), it supports dissipative Kerr solitons, and under normal dispersion, tailored excitation yields dark-pulse (``platicon'') states \cite{DelHaye2007,Herr2014a,Liu2018,Gaeta2019,Lobanov2015a}. Kerr effects, however, do not describe all observed dynamics. A second pathway is stimulated Raman scattering (SRS), which provides a broadband gain in microresonators. Raman lasers and Raman combs are established in fused silica and crystalline whispering-gallery devices \cite{Spillane2002,Liang2011,Farnesi2014,Kato2017,Andrianov2020}. They have also been demonstrated in CMOS-compatible materials including silicon, silicon carbide, and lithium niobate \cite{Griffith2016,Li2024,Yu2020,Lin2016}, and related AlN-on-sapphire work under self-injection locking (SIL) has shown stimulated Raman lasing and Stokes platicon microcombs and gigahertz-scale frequency sweeps \cite{Ding2025}. Theory has examined Raman-Kerr coupling, competition, and comb dynamics \cite{Milin2015,Chembo2015,Bao2015,Cherenkov2017,Okawachi2017,Choi2023,Liu2021,Andrianov2024,Andrianov2025}, and experiments in silica have shown Stokes-soliton formation with clear thresholds, intermodal coupling, and locking of repetition rates \cite{Yang2017a}. In Si$_3$N$_4$, by contrast, Raman effects were mainly observed as the self-frequency shift of Kerr solitons under anomalous dispersion \cite{Karpov2016}, while direct on-chip Raman frequency combs, especially coherent dark-pulse Raman states, have remained a key goal.

\begin{figure}[htbp!]
\centering
\includegraphics[width=\linewidth]{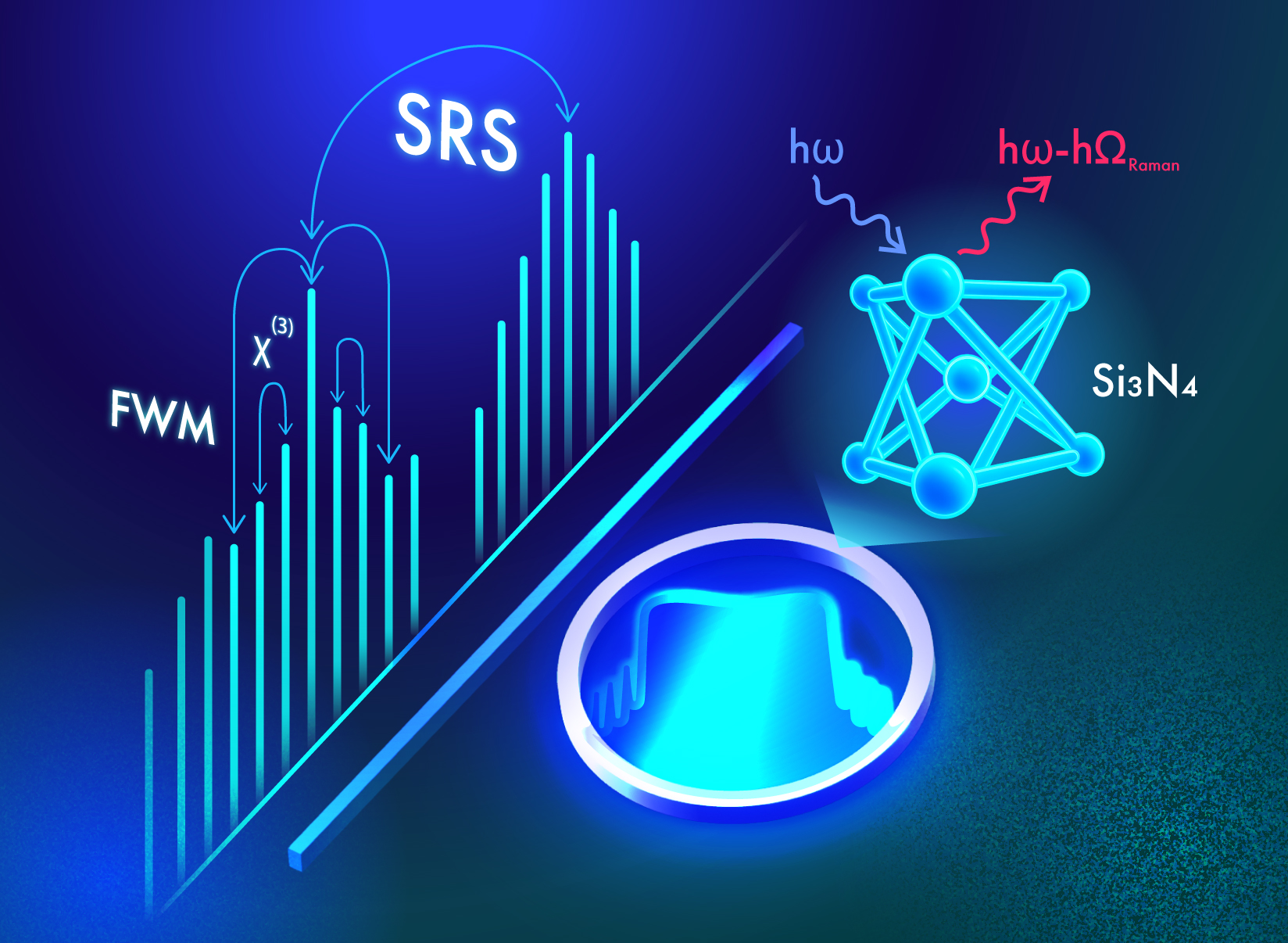}

\caption{The conceptual illustration of the process of Stokes frequency comb generation. The stimulated Raman scattering initiates four-wave mixing.  }
\label{fig:sketch}
\end{figure}

Here we experimentally demonstrate on-chip stimulated Stokes frequency combs in Si$_3$N$_4$ microring resonators, including dark-pulse Stokes states. Using ultra-high-$Q$ resonators and high intracavity power, we observe cascaded SRS with a frequency shift of $\simeq 9$~THz and a gain bandwidth of $\simeq 5$~THz. The resulting Stokes combs span $>$100~nm. To our knowledge within on-chip Si$_3$N$_4$ platforms, Stokes-comb generation, and in particular platicon-like Stokes comb states, has not been reported. From a broader perspective, these states exemplify non-equilibrium dissipative structures sustained by competing and cooperating nonlinear channels. The Raman pathway reshapes phase matching and modal coupling, offering a universal route to hybrid comb formation in driven–dissipative resonators. 

We realize these Stokes combs with two practical pump schemes: (i) a tunable laser with an isolator and (ii) a diode laser operated in the self-injection-locking regime, where coherent backscattering from the high-$Q$ cavity provides optical feedback for stabilization and linewidth narrowing \cite{Kondratiev2023, Kudelin2024}. 
In the SIL regime, we further show controlled switching between predominantly Kerr-comb and predominantly Stokes-comb operation by tuning the phase shift of the back-reflected wave and pump frequency detuning, which opens up new opportunities for practical applications. Numerical modeling with Raman gain reproduces our observations and indicates that SRS can seed and assist dark-pulse formation at both pump and Stokes frequencies via four-wave mixing pathways \cite{Milin2015,Okawachi2017,Liu2021}. These results point to a compact, CMOS-compatible route to hybrid Kerr-Raman comb sources in Si$_3$N$_4$ for coherent communications, low-noise microwave generation, and precision spectroscopy \cite{Marin-Palomo2017,Kudelin2024,Picqu2019,Suh2019}.

\begin{figure*}[htbp!]
\centering
\includegraphics[width=0.8\linewidth]{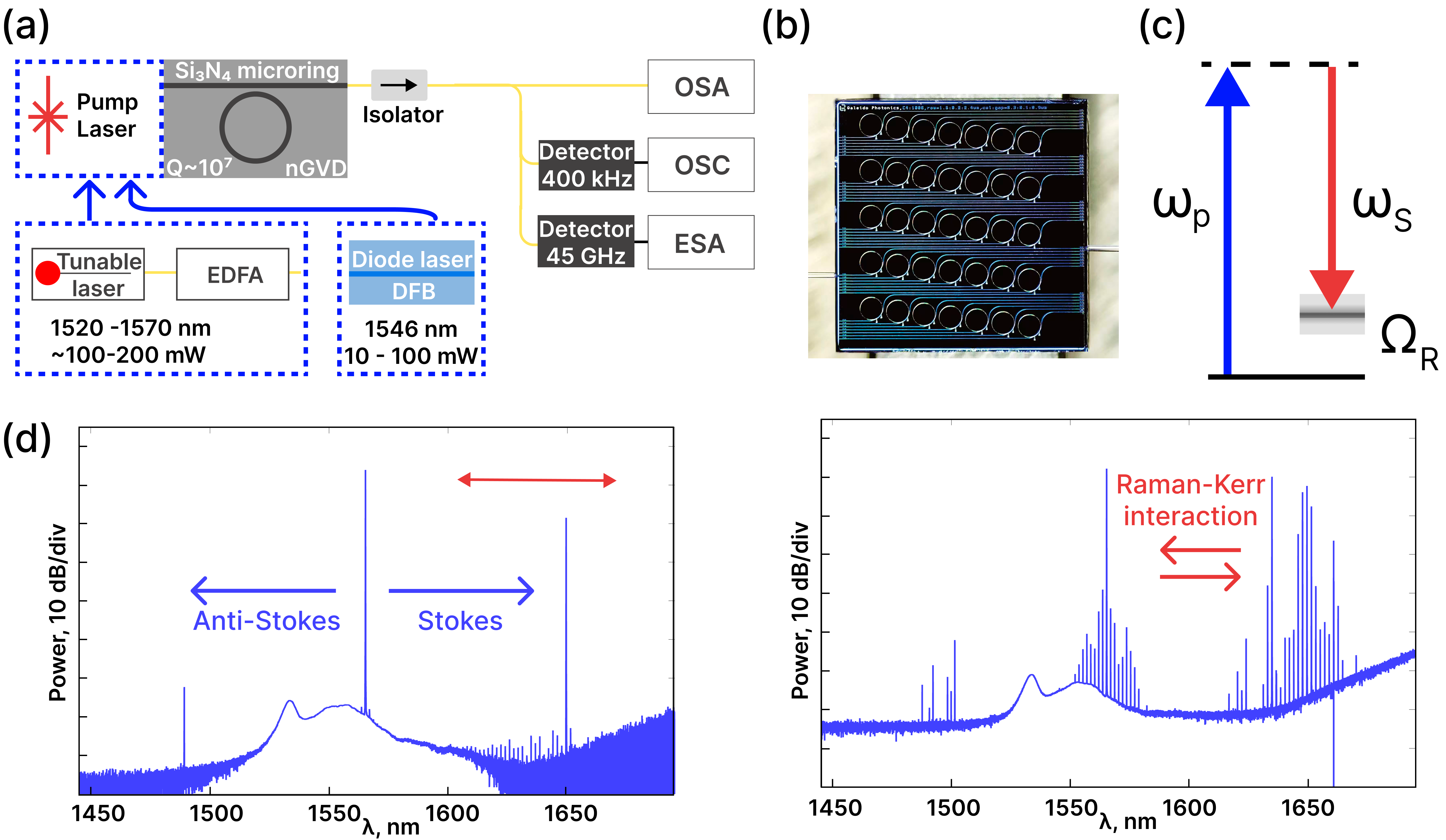}

\caption{(a) Experimental setup. Two types of laser sources are used as a pump: an amplified tunable laser with isolator and a DFB laser diode. A high-Q silicon nitride on-chip microresonator with normal GVD is investigated. An OSA, an ESA and an oscilloscope are used for measurements. (b) Photo of the chip. (c) Schematic energy diagram of the stimulated Raman scattering process, where $\omega_p$ and $\,\,\Omega_R$ are the pump and Raman frequencies, and $\,\, \omega_S \,\,$ is the Stokes shift. (d) Experimental spectra of Stokes generation. The left spectrum corresponds to the onset of generation, while the lower spectrum corresponds to higher coupled power into the microresonator.  }
\label{fig:setup}
\end{figure*}

\section{Experimental setup}

We conducted research using several high-Q microresonators with different gaps between the microresonator and the waveguide that set distinct coupled rates. The photo of the microresonators is presented in Fig.~\ref{fig:setup}(b) \cite{Ye2023}.
We present the results obtained for two representative microresonators: a critically coupled microresonator (the first one) and an undercoupled microresonator (the second one); see detailed characterization in the Supplementary Information (SI). Table ~\ref{tab:char} summarizes the key parameters of the system. We list the loaded quality factor Q, coupling efficiency $\eta$, free spectral range $D_{1}/2\pi$, and GVD coefficient $D_{2}/2\pi$. The values of $D_1/2\pi$ and $D_2/2\pi$ were determined using a wide-range frequency scan of the pump laser and frequency calibration with a Mach-Zehnder interferometer \cite{Dmitriev2022}. 
 The waveguide cross-section  is 300x2500 nm. The parameters for the critically coupled microresonator are used for numerical simulation.
\begin{table}[htbp!]
\centering
\caption{\bf Parameters of the microresonators}
\begin{tabular}{ccccc}
\hline
 & \,\,\,\,\,Q, $\cdot10^6$ \,\,\,\,\,\,&\,\,\,\,\,\,$\eta$ \,\,\,\,\,\,  &  $D_{1}/2\pi$, GHz\,\,  &  $D_{2}/2\pi$, MHz\,\,  \\
\hline
1 \,\, & $5$ & 0.5 & 202 & -39.4 \\
2 \,\,& $10$ & 0.25 & 184 & -32.6 \\

\hline
\end{tabular}
  \label{tab:char}
\end{table}

The experimental setup is shown in Fig. \ref{fig:setup}(a). We employ two types of pump sources. The first is a widely tunable (1510-1630 nm) single-frequency laser equipped with an optical isolator; its output signal is optically amplified and is coupled to the chip via a lensed fiber. 
The second is a single-frequency distributed feedback (DFB) laser diode emiting at 1546 nm, which is butt-coupled to a silicon nitride chip.   

\begin{figure*}[htbp!]
\centering
\includegraphics[width=0.8\linewidth]{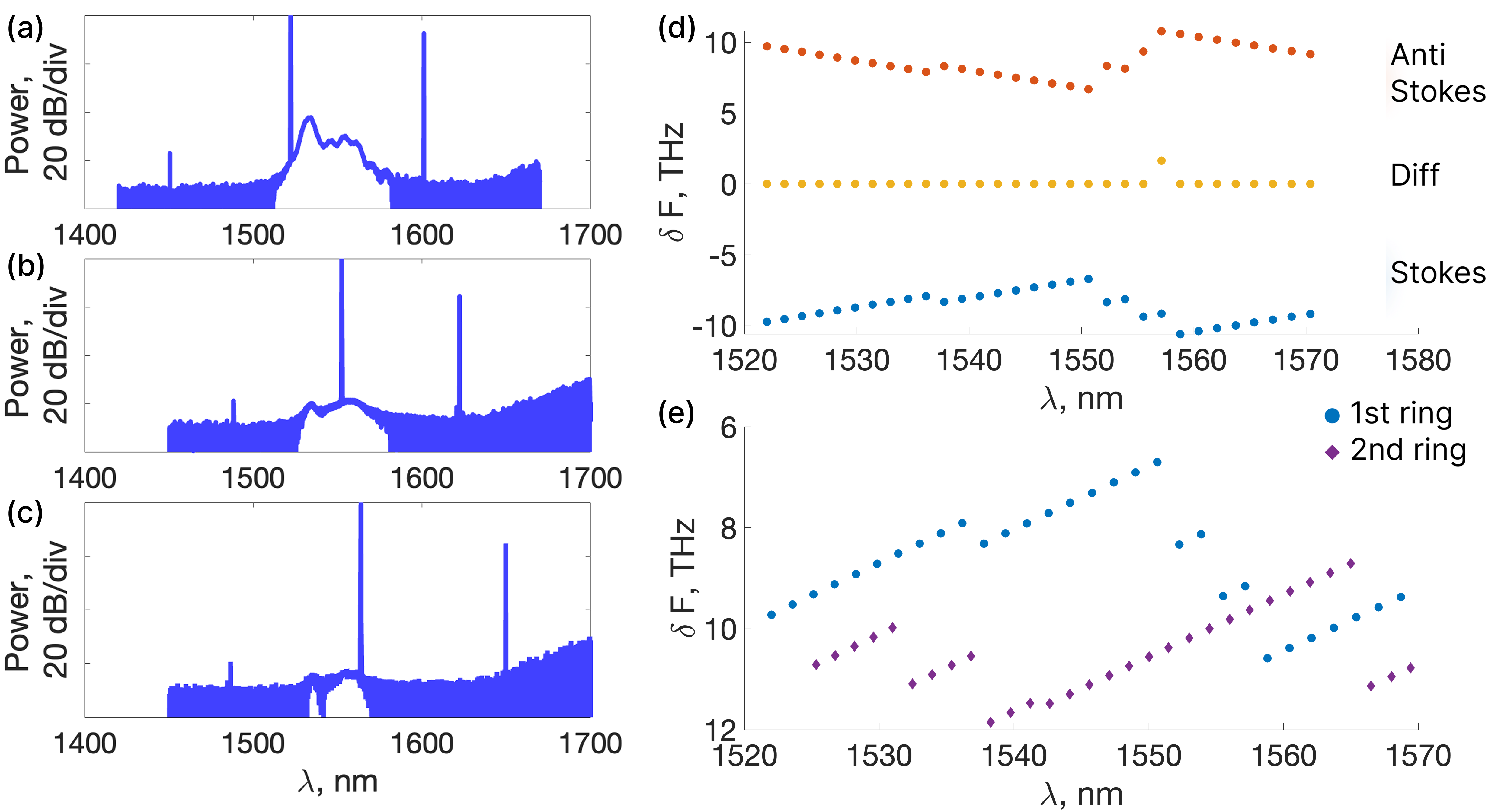}

\caption{Characterization of initial SRS. (a)-(c) Initial SRS spectra at different pump wavelength. (d)  Dependence of the Stokes and anti-Stokes frequency shifts (blue and red dots, respectively) on the pump wavelength in the critically coupled microring. Yellow dots represent the difference between these two frequency shifts, which appears to be equal within the resolution of the OSA. (e)  Dependence of the Stokes frequency shift on the pump wavelength for critically coupled and undercoupled microrings. The frequency shift is approximately 9 THz, and the Raman gain bandwidth is about 5 THz.  }
\label{fig:Raman_start_iso}
\end{figure*}

\begin{figure*}[htbp!]
\centering
\includegraphics[width=\linewidth]{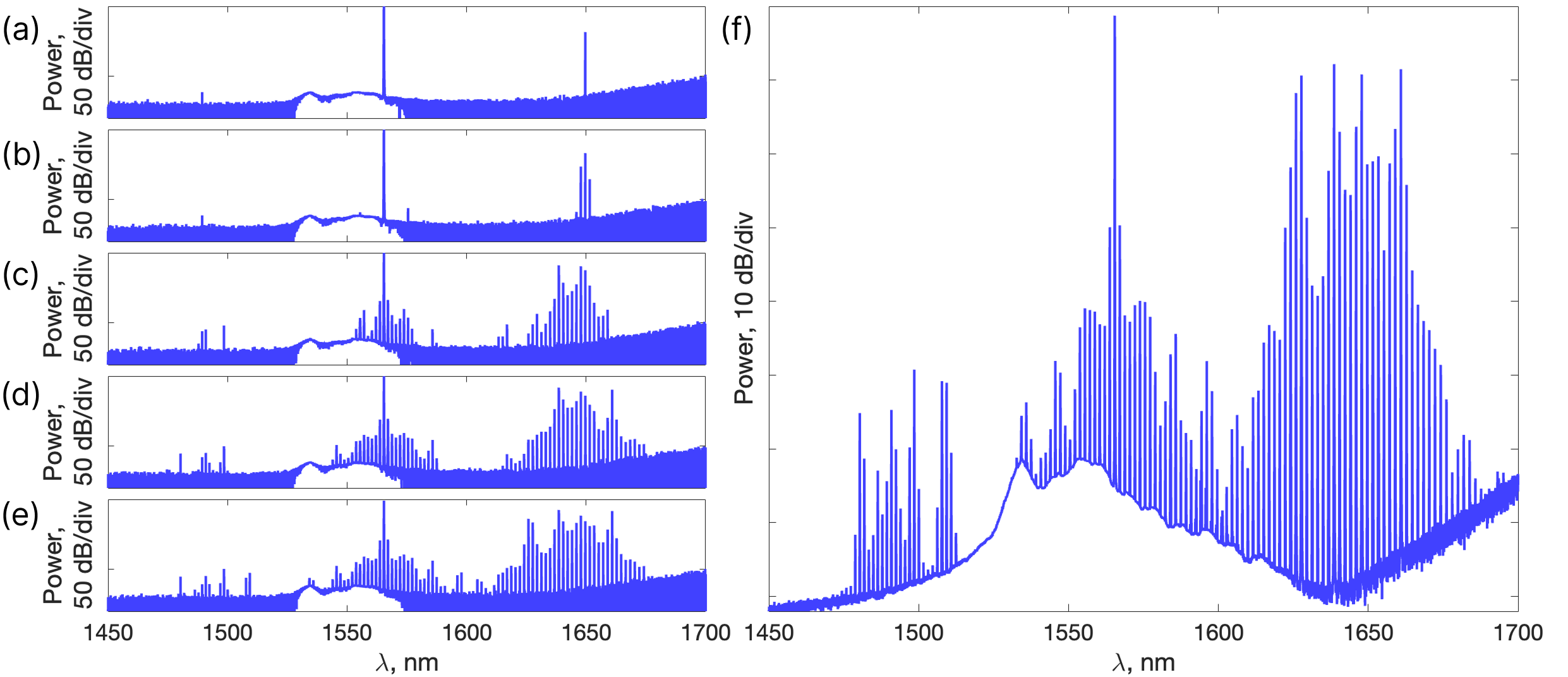}

\caption{Raman comb evolution with the increase of the intracavity power. (a)-(e) The spectra of the Raman comb generation with the increasing of the coupled to microresonator energy. The pump line power decreases by 6 dB from (a) to (e). (f) The spectrum obtained at pump power of 60 mW coupled into the chip. }
\label{fig:Raman_evo}
\end{figure*}

The chip output is coupled to a lensed single-mode optical fiber (SMF-28). For DFB-laser pumping, the on-chip power $P_{in}$ is $\approx$ 36\,mW at a 300\,mA operating current of the laser diode and $\approx$ 6\,mW at 100\,mA. For the tunable laser with an isolator, the pump power spans 30--90\,mW depending on the wavelength. An optical isolator after the output fiber suppresses the back-reflection into the chip. The transmitted signal is monitored on an oscilloscope (OSC), an electrical spectrum analyzer (ESA) and an optical spectrum analyzer (OSA). Also, a dedicated coupler allows injection of a reference laser for heterodyne beat-note detection on the ESA. The chip and laser diode are thermally stabilized with Peltier elements.

For the laser with an isolator, we gradually sweep the pump frequency through the resonance into red detuning. The excited frequency comb remains exceptionally stable due to the strong thermo-locking effect \cite{Carmon2004}.
With the DFB laser, we excite a microring mode by tuning the diode current, after which the laser enters the self-injection-locked (SIL) regime. 
By varying the diode–chip separation, we tune the locking phase and deterministically switch between predominantly Kerr and predominantly Stokes comb generation. The pump power is adjusted by the laser diode temperature with an appropriate current adjustment.

\begin{figure*}[htbp!]
\centering
\includegraphics[width=\linewidth]{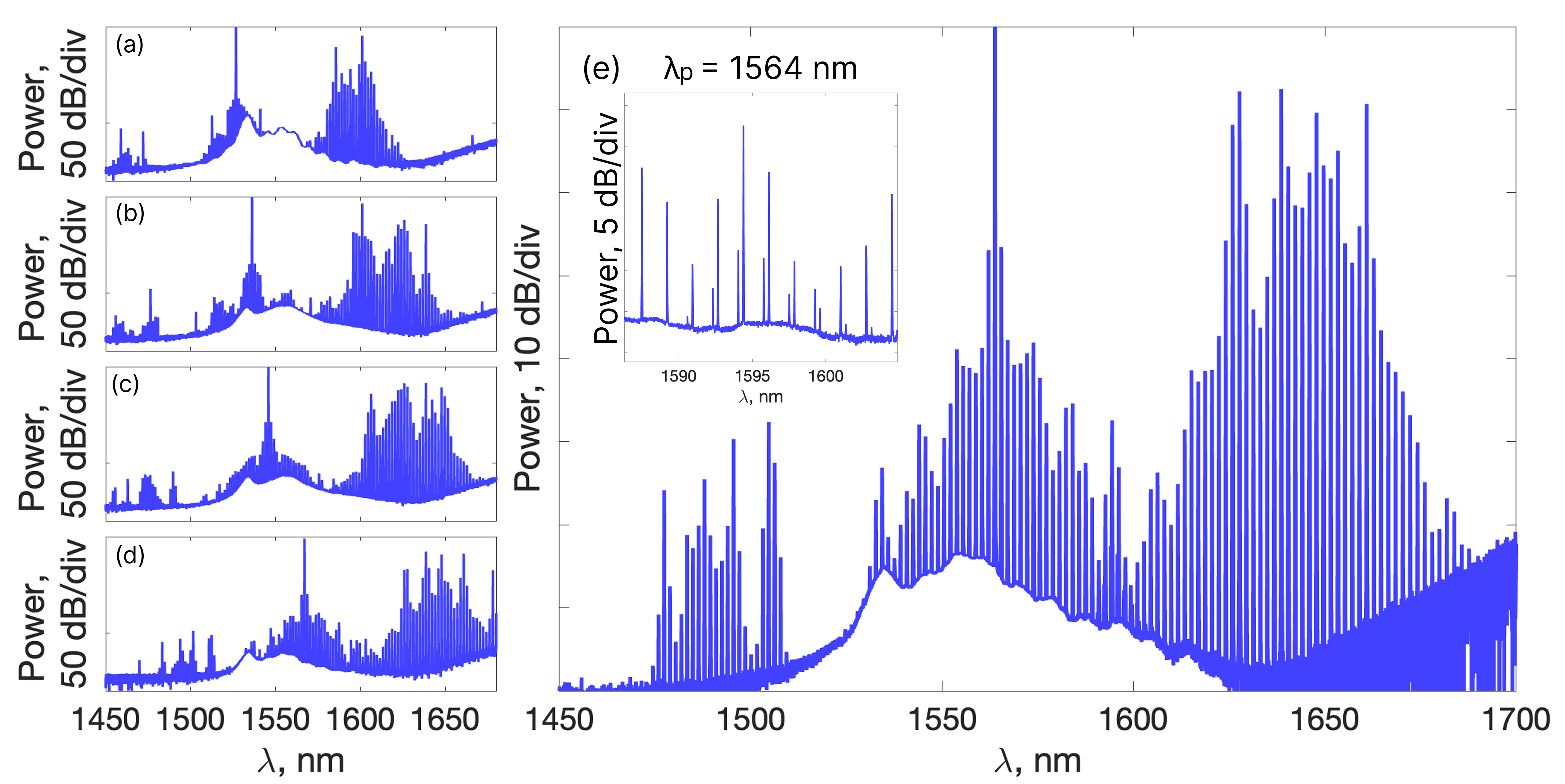}

\caption{The Stokes frequency comb obtained in case of pumping with stand alone amplified laser source. (a)-(e) The spectra obtained with pumping at 1526, 1536, 1545, 1567, and 1564 nm consequently are presented. Both Stokes and anti-Stokes components are observed. In the inset in panel (e) The sub-comb is observed near 1600 nm.}
\label{fig:Raman_TO}
\end{figure*}

\section{Initial Raman comb generation}

We pump microresonators in a wide wavelength range from 1520 to 1570\,nm with the tunable laser. The first Stokes line appears at different wavelengths from 1601 to 1663 nm depending on the pump wavelength, see Fig. \ref{fig:Raman_start_iso}.
The starting spectra for the different pump wavelengths for the critically coupled microring resonator are presented in Fig. \ref{fig:Raman_start_iso}(a)-(c). The Stokes component is accompanied by an anti-Stokes line at a shorter wavelength. Fig.~\ref{fig:Raman_start_iso}(d) plots the frequency shifts of both components relative to pump (red and blue) and their difference (yellow). 
The absolute values of frequency shifts are nearly equal with 0.04\,nm residual mismatch (except for one point). The OSA resolution is 0.02\,nm in these measurements. 
Fig. \ref{fig:Raman_start_iso}(e) shows the Raman frequency shift dependence on the pump wavelength for the critically coupled and under-coupled rings (blue and purple markers, respectively). The first Stokes line appears in the wide range of the frequency shift from 12 to 6\,THz. From these data one may estimate the Raman frequency shift as $\sim9$\,THz and Raman gain bandwidth as $\sim5$\,THz. This frequency shift of the Raman peak is consistent with values reported in Ref.~\cite{Mandal2024}, who attribute it to a longitudinal acoustic mode in Si nanocrystals embedded in silicon nitride \cite{Mankad2012}. \textcolor{black}{On the other hand, the frequency shift is close to measured in silica \cite{Spillane2002, Yang2017a} 
The initiation of Raman generation takes place at approximately 5-9~mW coupled to microresonator. \textcolor{black}{For comparison, \cite{Karpov2016} reported on-chip powers exceeding 3~W in a 100-GHz Si$_3$N$_4$ microresonator with $Q\approx10^6$, but no Stokes components were observed, which is presumably the result of Kerr comb generation in a thick (800 nm) waveguide with anomalous GVD, which accumulated power necessary for Stokes comb excitation.} The variation of the initial wavelength can be attributed to the amorphous structure of silicon nitride and cladding leading to wide gain width. 
Parameters of the modes at which the initial Stokes line appears and those adjacent to them are presented in Tab. \ref{tab:RModes}. The modes in which the first line appears are marked with ** sign. }

\begin{table}[htbp!]
\centering
\caption{\bf Parameters of the microresonator modes around Raman appearance wavelengths}
\begin{tabular}{ccccc}
\hline
 $\lambda$,& \,\,$\kappa_c$, \,\,\,\,\,\,&\,\, $\kappa_i$, \,\,\,\, & \,\,\,\,\,\,\,\,$\gamma$, \,\,\,\,\,\,\,\,  &  Q \,\,  \\
 nm & MHz & MHz &  MHz & $\cdot 10^6$ \\
\hline
1598.9 & 32.3 & 15.6 &  47 & 4.0   \\
1600.7**& 33.5 & 18.7 & 88 & 3.7  \\
1602.4&  34.7 & 18.1 &  60 & 3.6  \\
1604.1&  30.0 & 15.0 & 23 & 4.3  \\
1605.9**&  25.0 & 18.0 &  30 & 4.5  \\
1607.6&  36.0 & 18.0 &  51 & 3.6  \\
1609.4&  30.7 & 12.8 &  44 & 4.5  \\
1611.2&  31.5 & 14.2 &  22 & 4.2  \\

\hline
\end{tabular}
  \label{tab:RModes}
\end{table}

\section{Cascaded Stokes microcomb generation}

We excite a frequency comb by continuously scanning the pump frequency using an amplified laser with an optical isolator. The resonance curves are nonlinear because of thermo-optical and Kerr effect at high pump power. As the pump frequency is decreased, we move along the resonance curve toward resonance and increase the intracavity power. 
In Fig.~\ref{fig:Raman_evo}(a)-(f) the evolution of the spectra during such scan is presented. For these spectral measurements, the coupled on-chip power is $P_{in}\approx60$ mW. The first Stokes line appears at incoupled into microresonator power $P_{m}\approx14$\,mW, see Fig.~\ref{fig:Raman_evo}(a). Then, the lines around the initial one appear along with sidebands around the pump line, see Fig.~\ref{fig:Raman_evo}(b). Further increasing of the coupled power ($P_m\approx25$\,mW) leads to formation of the cascaded Stokes microcomb, see Fig.~\ref{fig:Raman_evo}(c). Notably, the strongest Raman line is shifted to the red side from the first excited one. A Kerr comb also appears around the pump but its power is about 20\,dB lower than the Stokes comb. The Raman comb continues to broaden (see Fig.~\ref{fig:Raman_evo}(d) and (e)) while the power of the initial frequency lines remains almost the same.  
Figure~\ref{fig:Raman_evo}(f) shows the spectrum at the maximum red detuning at which the Raman comb is generated. The Kerr comb and Raman comb overlap at $\approx$1600\,nm. 


To investigate how Raman comb generation depends on the pumped microresonator mode, we varied the pump wavelength over a broad range. The measured spectra are shown in Fig.~\ref{fig:Raman_TO}. Figs.~\ref{fig:Raman_TO}(a)-(d) present the spectra for the largest observed pump detunings for pumped modes at 1526, 1536, 1545, 1567\,nm. The Stokes frequency comb shifts to longer wavelength following the pump. The on-chip pump power varies with wavelength ($P_{in} = $ 47.7, 64.5, 79.2, 85.5\,mW, respectively) owing to the wavelength-dependent performance of the optical amplifier. A distinct line at 1670\,nm, visible in Figs.~\ref{fig:Raman_TO}(a) and (b), lies noticeable apart from the Stokes frequency cluster and likely produced by four-wave mixing between the pump and the Stokes line. 
The Stokes comb lines in Fig. \ref{fig:Raman_TO} exhibit power comparable to that of the pump, with the strongest lines only 5-8 dB lower. In Fig.~\ref{fig:Raman_TO}(b), the Stokes line is notably effective -- approximately 5~dB below the pump. 
In contrast, the generation of the Kerr platicon at the pump wavelength is much less efficient. Its lines are more than 30~dB lower than the pump and more than 20 dB lower than the Stokes comb lines. 
In Fig.~\ref{fig:Raman_TO}(e), the spectrum of the comb pumped at 1564\,nm is presented. The inset presents the overlap between the Kerr and Stokes combs area. 

\begin{figure}[htbp!]
\centering
\includegraphics[width=\linewidth]{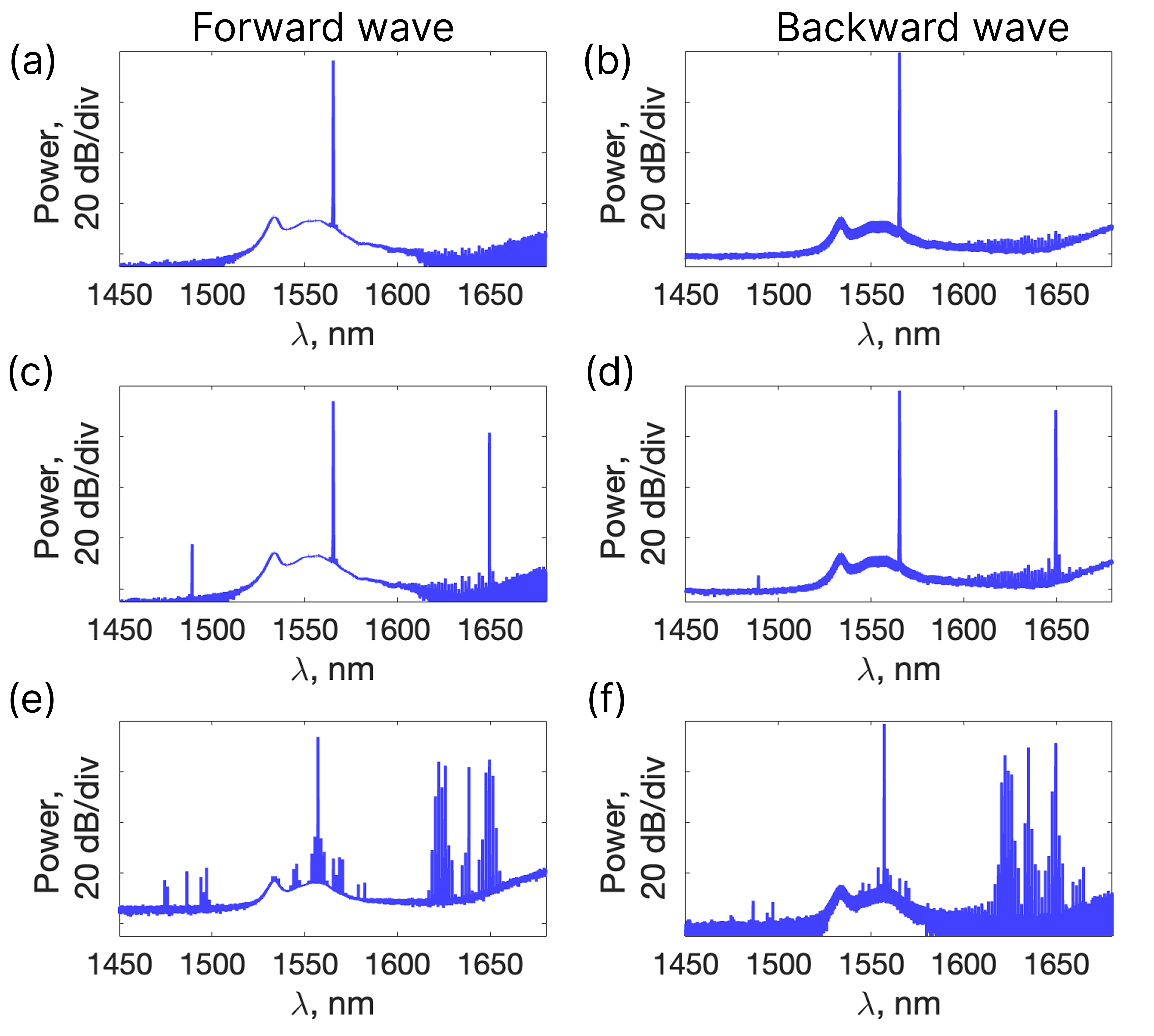}

\caption{The spectra of the combs in forward (left column) and backward (right column) directions. The spectra in a row are taken simultaneously. The lower the row the higher the detuning. One may see, that the backward wave from the pump line is almost constant, while the forward pump line decreases for more than 4 dB due to the coupling.}
\label{fig:Raman-backward}
\end{figure}

Fig. \ref{fig:Raman-backward} compares the forward wave and backward wave spectra, recorded simultaneously using two OSAs. The first column shows the forward spectra and the second one is the backward spectra. The rows are ordered by increasing on-chip coupled power. Weak lines at 1600-1650\,nm range indicate the Raman gain region [panels (b, d)]. Comparing the forward and backward Raman spectra [panels (e) and (f)] reveals different spectral envelopes, indicating a strong wavelength dependence of the backscattering coefficients.
\textcolor{black}{The pump line power in spectra of the forward wave drops by more than 4 dB (from -3.7 to -7.9 dBm) between the first and third rows [panels (a), (c), (e)], while the backward pump line power changes noticeably less, within 1 dB (from -0.5 to -1.1 dBm) [panels (b), (d), (f)]. The first Stokes-line has almost constant line power within increasing power while new lines appear. It is evident that the comb around the pumping is weaker in the backward wave, [see Fig. \ref{fig:Raman-backward} (e, f).] }
We also measured the forward and backward waves in case of cascaded Stokes comb generation separately for both TE and TM polarizations. No signal was detected in orthogonal (TM) polarization.

\section{Numerical model}

To numerically investigate the mechanism of Stokes microcomb generation in Si$_3$N$_4$ microring resonators, we employ a system of coupled mode equations (CME) \cite{Yang2017} in normalized form (see Supplemental Information for details). Such a system describes the temporal evolution of the field amplitudes of forward and backward waves inside microresonator at both pump and Stokes frequencies:

\begin{widetext}
    \begin{multline}
       \label{eq:CME} 
   \frac{\partial a_{\mu}}{\partial\tau} = -[1 - i \zeta_{\mu} ] a_\mu + i\beta b_{\mu}  + \delta_{0,\mu} f + i \mathcal{F}^{-1}([|\tilde a_{\mu}|^{2} - \tau_r  D_1 f_r \frac{\partial (|\tilde a_{\mu}|^{2} + |\tilde{a}^{(r)}_{\mu}|^{2})}{\partial \varphi} + (2 - f_{r}) |\tilde{a}_\mu^{(r)}|^2] \tilde{a}_{\mu})+ \\
    + i (2 - f_{r}) (\sum_\nu(|b_\nu|^2 +|b^{(r)}_\nu|^2))a_{\mu}
    -  \frac{\omega_0}{\omega_r} G_{r}\sum_{\nu}(|{a}^{(r)}_{\nu}|^2 + |b^{(r)}_{\nu}|^2)a_\mu   ,
    \end{multline}

\begin{multline}\label{eq:CME_back}
   \frac{\partial b_{\mu}}{\partial\tau} = -[1 - i \zeta_{\mu}] b_\mu + i\beta a_{\mu}  +  i \mathcal{F}([|\hat b_{\mu}|^{2} - \tau_r  D_1 f_r \frac{\partial (|\hat b_{\mu}|^{2} + |\hat{b}^{(r)}_{\mu}|^{2})}{\partial \varphi} + (2 - f_{r}) |\hat{b}_\mu^{(r)}|^2] \hat{b}_{\mu}) +\\
    + i (2 - f_{r}) (\sum_\nu(|a_\nu|^2 +|a^{(r)}_\nu|^2))b_{\mu} 
    - \frac{\omega_0}{\omega_r} G_{r}\sum_{\nu}(|{b}^{(r)}_{\nu}|^2 + |a^{(r)}_{\nu}|^2)b_\mu ,
 \end{multline}

\begin{multline}\label{eq:CME_raman}
    \frac{\partial a^{(r)}_{\mu}}{\partial\tau} = -[1 - i \zeta_{\mu} -\Delta] a^{(r)}_\mu + i\beta b^{(r)}_{\mu} + i \mathcal{F}^{-1}([|\tilde a^{(r)}_{\mu}|^{2} - \tau_r  D_1 f_r \frac{\partial (|\tilde {a}^{(r)}_{\mu}|^{2} + |\tilde{a}_{\mu}|^{2})}{\partial \varphi} + (2 - f_{r}) |\tilde{a}_\mu|^2] \tilde{a}_{\mu}^{(r)}) + \\
    + i (2 - f_{r}) (\sum_\nu(|b^{(r)}_\nu|^2 +|b_\nu|^2))a^{(r)}_{\mu} 
    +  G_{r}\sum_{\nu}(|{a}_{\nu}|^2 + |b_{\nu}|^2)a^{(r)}_\mu ,
 \end{multline}

\begin{multline}\label{eq:CME_raman_back}
   \frac{\partial b^{(r)}_{\mu}}{\partial\tau} = -[1 - i \zeta_{\mu} -\Delta] b^{(r)}_\mu + i\beta a^{(r)}_{\mu} + i \mathcal{F}([|\hat b^{(r)}_{\mu}|^{2} - \tau_r  D_1 f_r \frac{\partial (|\hat {b}^{(r)}_{\mu}|^{2} + |\hat{b}_{\mu}|^{2})}{\partial \varphi} + (2 - f_{r}) |\hat{b}_\mu|^2] \hat{b}_{\mu}^{(r)})+ \\
   + i (2 - f_{r}) (\sum_\nu(|a^{(r)}_\nu|^2 +|a_\nu|^2))b^{(r)}_{\mu} + G_{r}\sum_{\nu}(|{b}_{\nu}|^2 + |a_{\nu}|^2)b^{(r)}_\mu,  
   \end{multline}
\end{widetext}
where $a_\mu$ and $b_\mu$ denote the field amplitudes of the forward- and backward- propagating modes of the microresonator with mode index $\mu$, respectively; $\tau$ is the normalized time ($\tau = 2t/\kappa_m$), $\kappa_m = \kappa_0 + \kappa_c$ is the microresonator total decay rate including intristic losses ($\kappa_0$) and external coupling rates ($\kappa_c$), $\zeta_\mu$ represents the normalized frequency detuning from an FSR-equidistant frequency grid, $\Delta$ is a normalized difference between the FSRs near the pumped mode and Stokes-mode frequencies,  $\beta$ is the normalized coupling rate between counter-propagating modes inside the microresonator, $\delta_{0,\mu}$ is the Kronecker delta, $f = \sqrt{\frac{8 \eta \hbar \omega_{0}^2 c n_2}{\kappa_m^2n_0^2 V_\text{eff}}} \sqrt{\frac{P_{\text{in}}}{\hbar \omega_{0}}}$ is the dimensionless pump amplitude, $\eta = \kappa_c/\kappa_m$ is the coupling efficiency ($\eta = 0.5$ corresponding to critical coupling, $\eta= 1$ to strong overcoupling), $\omega_0$ is the pump frequency, $P_{\text{in}}$ are the input pump power, $n_0$ is the refractive index, $n_2$ is the nonlinear refractive index, $V_\text{eff}$ is the effective mode volume, $c$ is the speed of light in vacuum, $\hbar$ is the reduced Planck constant, $\delta$ is a difference between repetition rates of the primary Kerr and Stokes solitons; $\mathcal{F}$ and $\mathcal{F}^{-1}$ denote the discrete and inverse discrete Fourier transforms,  respectively, $\tilde{ }$\, indicates the discrete Fourier transform, and $\hat{ }$\, indicates the inverse discrete Fourier transform, $\varphi$ is the angular coordinate along the microresonator, $f_r$ is the Raman fraction, $\tau_r$ is the normalized Raman shock time, $D_1$ is the free-spectral range, $G_r$ is the normalized Raman gain, and the superscript $^{(r)}$ denotes Stokes-wave amplitudes. 

   \begin{figure*}[htbp!]
\centering
\includegraphics[width=0.96\linewidth]{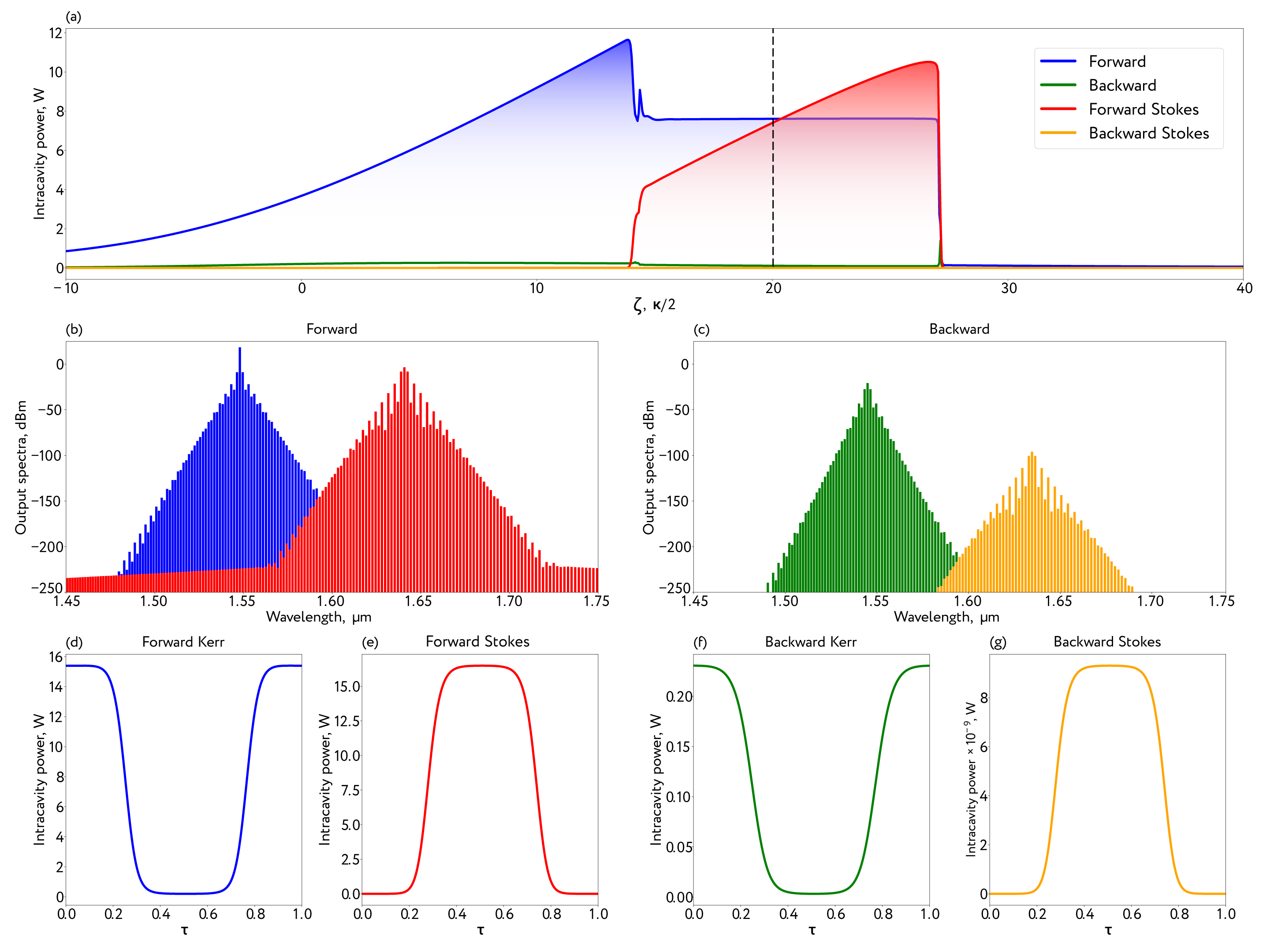}

\caption{(a) Intracavity power of the forward and backward waves near the pump and Stokes frequencies (blue, green, red, and yellow lines, respectively). The platicon step occurs synchronously with the onset of the Stokes components. (b,c) Optical spectra of the forward (b) and backward (c) waves at a detuning of 20 (dashed line in panel (a)). (d–g) Platicon profiles corresponding to the spectra shown above: (d) forward-wave pump, (e) forward-wave Stokes, (f) backward-wave pump, (g) backward-wave Stokes.}
\label{fig:num}
\end{figure*}

\begin{figure*}[htbp!]
\centering
\includegraphics[width=0.98\linewidth]{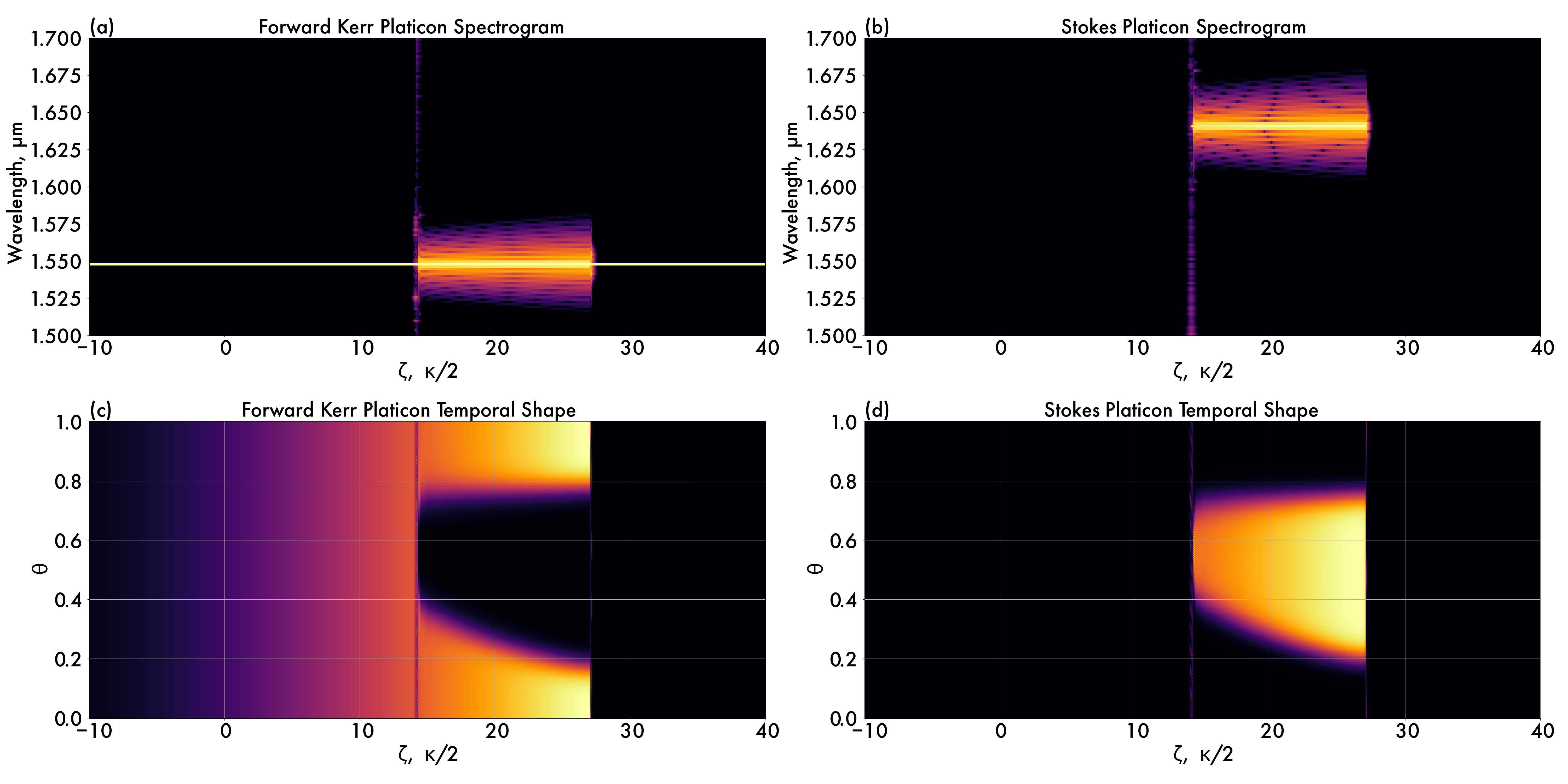}

\caption{Evolution of the spectra (a,b) and temporal profiles (c,d) of forward-propagating Kerr and Stokes platicons, respectively, during a pump-frequency scan.}
\label{fig:num2}
\end{figure*}

    We account for both forward and backward waves at the pump and Stokes frequencies. The numerical simulations use parameters obtained directly from the experiment.
    Fig.~\ref{fig:num} summarizes the simulation results. At a certain detuning, platicon generation begins simultaneously at both the pump and Stokes frequencies, as seen in the intracavity power traces in Fig.~\ref{fig:num}(a). The power of Stokes components is higher than the Kerr platicon comb lines [Fig.~\ref{fig:num}(a, b)]. The corresponding platicon temporal profiles for point $\zeta_{\mu} = 20$.
    Interestingly, the platicon  profiles at the pump and Stokes wavelengths alternate in time, see Fig.~\ref{fig:num2} (c) and (d). The numerical simulations show that the generation of coherent frequency combs in normal GVD microresonator, in the presence of the Raman scattering, can occur at both pump and Stokes wavelengths. Furthermore, no Kerr-comb generation was observed during frequency scanning in the absence of Raman gain, highlighting the crucial role of the Stokes component in initiating the comb through nonlinear coupling. 

\section{Stokes combs by Self-injection locked laser}

\begin{figure*}[htbp!]
\centering
\includegraphics[width=0.7\linewidth]{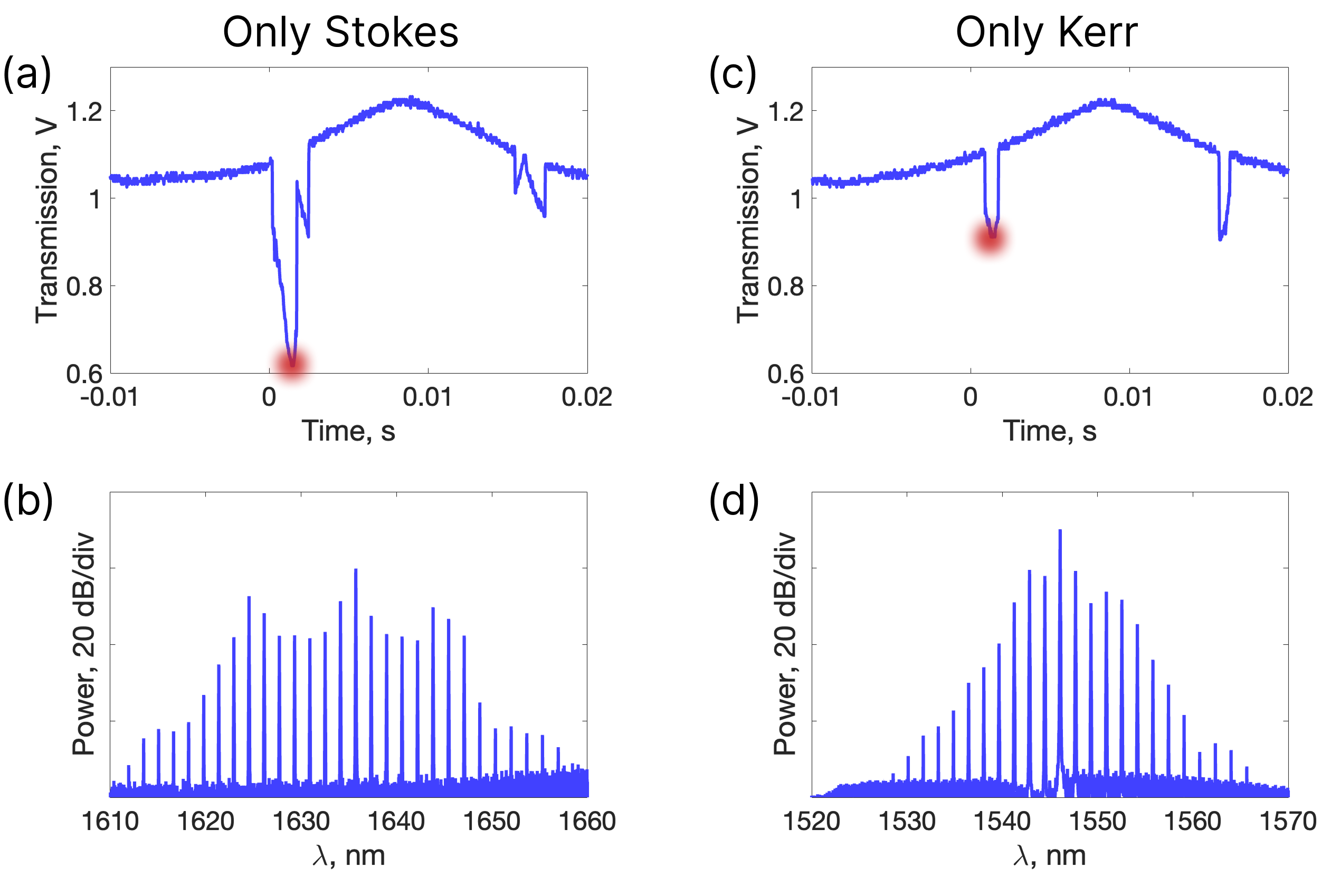}

\caption{Generation of Kerr comb or Stokes comb in case of SIL. Tuning the locking phase the necessary for Kerr comb generation detuning can be achieved in SIL regime. Varying the locking phase the generation of only Stokes comb or only Kerr comb generation can be achieved.}
\label{fig:Raman-Kerr}
\end{figure*}

Self-injection locking of semiconductor lasers to high-$Q$ microresonators has emerged as a powerful approach for compact and low-noise frequency-comb generation \cite{Raja2019, Liu2021a, Shen2020} and may be used for comb generation at normal GVD \cite{Jin2021, Li2025, Sun2025}. In this regime, coherent Rayleigh backscattering provides optical feedback that locks the laser emission to the cavity resonance, enabling ultranarrow linewidths and stable detuning control. Here we explore Stokes microcomb formation in the SIL regime and reveal how the locking phase determines whether Kerr or Stokes combs dominate. 


We find that, in contrast to the above results, SIL enables Kerr-comb generation even under normal GVD. In the SIL regime, the system can be switched between efficient Kerr-comb generation and purely Stokes-comb generation. The operating state is governed by the locking phase, determined by the optical path length between the laser diode and the microring, which can be precisely adjusted using a piezoelectric actuator.


Figure~\ref{fig:Raman-Kerr}(a) shows the transmission spectrum. The red marker indicates the point where Stokes frequency combs are observed [Fig.~\ref{fig:Raman-Kerr}(b)]. By varying the locking phase, we obtain less power coupled into the resonance and a corresponding change in detuning; under these conditions, a Kerr comb appears [Fig.~\ref{fig:Raman-Kerr}(c,d)]. The spectra in panels (b) and (d) can thus be switched by tuning the locking phase.


Thus, precise adjustment of the locking phase determines which comb regime is realized. As we proved in the numerical study, the coherent regime of the Stokes comb may appear. In Fig. \ref{fig:Raman-Kerr} (b) the shape of the Stokes comb has platicon-like envelope and the were no exsessive noise at the ESA close to zero frequency indicating coherent regime. 

In the SIL regime, the on-chip coupled power $P_{in}$ ranged from 6 mW to 50\,mW, which is sufficient for Stokes-comb generation according to the above results. However, even at high $P_{in}$, the Stokes comb generation can be avoided by choosing certain locking phase. So that, we make the conclusion, that not only enough field intensity is the key ingredient for the Stokes comb observation. At certain detunings, the Kerr comb dominates the intracavity dynamics, depleting the pump and preventing the Raman gain from reaching threshold. 
Conversly, at detunings that not support Kerr-comb formation, the Stokes comb appears. In case, when using a laser with an optical isolator, Stokes-comb generation occurs once the power is sufficient, because Kerr combs cannot be generated by simple frequency scan under normal GVD, and hence no competition arises between these two processes. 

These observations provide a basis for discussing the mechanism of Kerr-comb generation when pumping a normal-GVD microresonator with an isolated laser during a gradual frequency scan. We find that the emergence of a cascaded Stokes comb fundamentally alters the expected dynamics. Once the Stokes comb forms, it establishes conditions that enable Kerr-comb generation, although the Kerr comb remains approximately 10 dB weaker than the Stokes one. In contrast, when pumping with a SIL laser, conventional Kerr-comb formation becomes possible. If the pump detuning reaches a certain value, the Kerr comb forms and depletes the pump, thereby preventing Stokes-comb generation. Conversely, by adjusting the locking phase, the effective detuning can be shifted so as to suppress Kerr-comb formation and leave sufficient intracavity power for Raman lasing.



\begin{figure*}[htbp!]
\centering
\includegraphics[width=0.88\linewidth]{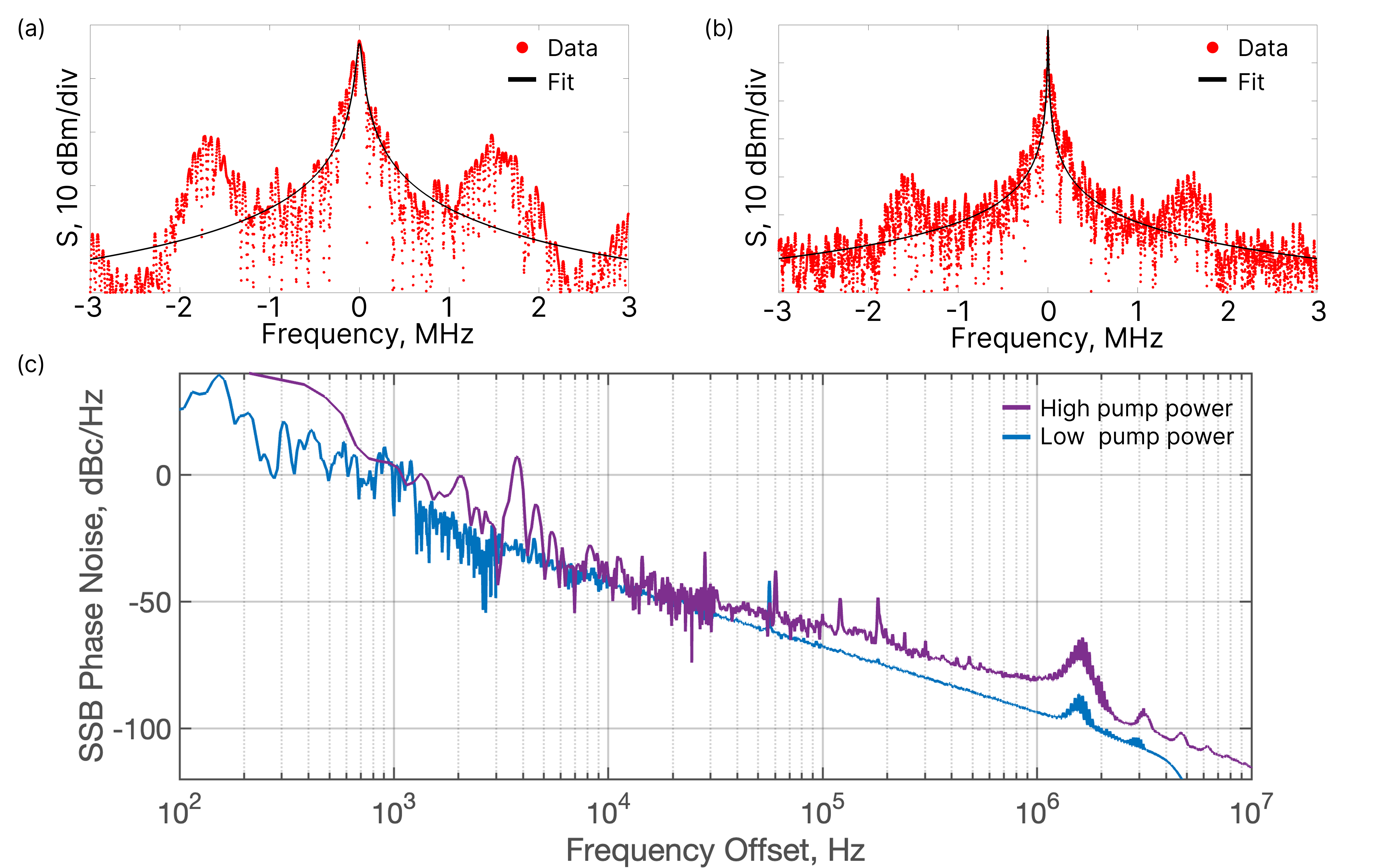}

\caption{Spectral characteristics of the pump in SIL regime in case of Stokes generation (left column), and linear regime (right column). The laser linewidth narrows by 10 times.}
\label{fig:spec_prop}
\end{figure*}



\section{influence of Raman scattering on SIL efficiency}

We monitored the spectral properties of the pump in the SIL regime by heterodying it with a reference laser. The linewidth turns out to be unexpectedly noisy for such a high-Q microring, see Fig. \ref{fig:spec_prop} (a), (b). The laser linewidth reached 30-60 kHz, as evident from both the beatnote measurements and the single-sideband phase-noise spectral density. The excess noise arises from nonlinear processes in the microresonator and increases further when a Stokes comb is generated. 

To verify that the nonlinear effects are responsible for the observed excessive noise, we reduced the pump power by more than 6 times, approaching the lasing-threshold regime. As the pump current was lowered, the laser diode temperature was simultaneously adjusted to maintain the pump wavelength. 
We perform this procedure at all mentioned microrings and achieve the regimes where no nonlinear generation has been observed. The measurements results are presented in Fig. \ref{fig:spec_prop}. Under these conditions, the diode laser linewidth narrows to approximately 6 kHz, extracted from a Lorentzian profile approximation. Phase noise also decreases in comparison to the case of a higher pump power. For every microresonator examined,lower laser diode current consistently results in reduced phase noise [Fig. \ref{fig:spec_prop}]. 

\section{Discussion}

Stimulated Raman scattering in silicon nitride photonic circuits establishes a broadly applicable platform for integrated nonlinear optics. The robust generation of the Stokes wave enables on-chip Raman lasing, where the intrinsic $\sim 9$-THz frequency shift grants access to spectral regions beyond the pump. Compared with previously demonstrated integrated Raman-lasing systems \cite{Yu2020, Xia2022, Zhang2021, Liu2017}, the approach reported here combines the versatility, CMOS compatibility, and low-loss performance of Si$_3$N$_4$ with strong conversion efficiency.
During the preparation of this manuscript, a preprint reporting Raman lasing in Si$_3$N$_4$ microrings appeared \cite{Zheng2025}. Although those results were obtained using chips fabricated by a different foundry, they exhibit remarkable qualitative and quantitative agreement with our observations—including normal GVD, high Q, and a discrete Stokes shift of $\sim 10$\,THz, highlighting the generality and robustness of Raman lasing in this platform.

Beyond single-line lasing, the evolution of the Stokes wave into a coherent frequency comb opens opportunities in high-resolution molecular spectroscopy, including detection of methane absorption features beyond 1600\,nm \cite{Coburn2018}. The broadband nature of the Raman comb is also promising for coherent optical communications, providing a compact source for wavelength-division multiplexing in the long-wavelength infrared \cite{Marin-Palomo2020}. Importantly, the Raman-comb lines exhibit high output power: in Fig.~\ref{fig:Raman_TO}(e), nine lines exceed 100\,\textmu W and two exceed 800\,\textmu W at the output fiber. The use of commercial, electrically driven laser diodes as pumps enhances practicality and strongly supports the technological readiness of this approach. Furthermore, the demonstrated ability to switch between Kerr- and Raman-based combs on the same chip—controlled solely by the locking phase in the SIL regime—provides a powerful degree of reconfigurability for advanced photonic systems, including dynamically tunable communication links and resilient quantum key-distribution networks.



To conclude, we have experimentally observed stimulated Raman scattering in silicon nitride microresonators operating under normal group-velocity dispersion. We report Raman lines with peak power exceeding –5 dBc relative to the pump and a Raman shift of $\sim$9\,THz with a gain bandwidth of $\sim$5\,THz. By simple frequency tuning of a laser equipped with an isolator, we generate hybrid frequency combs containing both Raman and Kerr components. The Stokes comb spans more than 100\,nm and coexists with a platicon-like Kerr comb, exhibiting strong thermo-locking stability. Using a self-injection-locked laser diode, we observe analogous behavior and demonstrate control over the comb regime by varying the locking phase. Numerical simulations corroborate the experimental results, revealing that coherent combs can arise simultaneously at pump and Stokes frequencies.

Our results uncover an intricate interplay between Kerr and Raman nonlinearities: the dominant process is governed by the pump-frequency detuning. In a normal-GVD resonator with dimensionless pump amplitude $f \in [3.5, 12]$, Raman generation is favored due to the suppression of Kerr four-wave mixing. Once the Stokes comb emerges, it creates favorable conditions for subsequent Kerr-comb formation at significantly lower power—approximately 10\,dB below the Raman lines. In contrast, within the SIL regime, Kerr four-wave mixing becomes accessible; when the detuning reaches appropriate values, Kerr processes redirect pump energy, thereby suppressing Raman lasing. Adjusting the locking phase thus toggles the system between Raman-dominated and Kerr-dominated operation.

Finally, we show that Stokes generation plays a major role in the SIL dynamics, substantially broadening the laser linewidth, an effect which can be mitigated by lowering the pump power. Altogether, these findings advance the fundamental understanding of nonlinear interactions in Si$_3$N$_4$ microresonators and open new avenues for Raman-based sources and reconfigurable photonic functionalities.

\eject
\bibliography{export.bib}
\end{document}